\documentclass[3p,times,twocolumn]{elsarticle}

\usepackage{ecrc}

\volume{00}

\firstpage{1}

\journalname{Nuclear Physics B Proceedings Supplement}

\runauth{S.~Antusch}

\jid{nuphbp}

\jnltitlelogo{Nuclear Physics B Proceedings Supplement}

\usepackage{amssymb}

\usepackage[figuresright]{rotating}
\usepackage[latin1]{inputenc}
\usepackage{dsfont}
\usepackage{amsfonts}
\usepackage{amsmath}
\usepackage{amssymb}
\usepackage{amsthm}
\usepackage{graphicx}
\usepackage{booktabs}
\allowdisplaybreaks[1]
\usepackage{color}
\usepackage{ulem}
\usepackage{tikz}
\usetikzlibrary{arrows}

\begin{document}

\newcommand{\be}{\begin{equation}}
\newcommand{\ee}{\end{equation}}
\newcommand{\beq}{\begin{equation}}
\newcommand{\eeq}{\end{equation}}
\newcommand{\bea}{\begin{eqnarray}}
\newcommand{\eea}{\end{eqnarray}}

\begin{frontmatter}



\dochead{}

\title{Models for Neutrino Masses and Mixings}

\author[label1,label2]{Stefan Antusch}
\address[label1]{ Department of Physics, University of Basel, Switzerland}
\address[label2]{Max-Planck-Institut f\"ur Physik (Werner-Heisenberg-Institut), M\"unchen, Germany}

\begin{abstract}
We review recent developments towards models for neutrino masses and mixings. 
\end{abstract}

\end{frontmatter}


\section{Introduction}
\label{sec:introduction}

The last year has been an exciting time for models of neutrino masses and mixings, especially due to the recent measurement of the leptonic mixing angle $\theta^\mathrm{PMNS}_{13}$ by T2K \cite{t2k}, DoubleCHOOZ \cite{Abe:2011fz}, DayaBay \cite{An:2012eh} and RENO \cite{Ahn:2012nd}, featuring
\be
\theta^{\text{PMNS}}_{13} = 8.8^\circ \pm 1.0^\circ  \:.
\ee
This rather accurate measurement has dramatic consequences for model building. Out of the many models proposed, a large fraction is now excluded and only a small fraction remains which can explain the experimentally found value. 

One might now think that theorists get depressed by so many models being ruled out, however of course just the opposite is true: The measurement has triggered an enormous interest in the community and a large number of theory papers appeared, analysing its consequences. 

Another reason why this field is so lively is the fact that there are still many unknowns in the neutrino sector, which means that models still have the possibility to make predictions. In this respect, the measured value of $\theta^\mathrm{PMNS}_{13}$ can now be used as an input parameter for building models to predict these currently unknow parameters, such as, e.g.,
\begin{itemize}
\item the Dirac CP phase $\delta^\text{PMNS}$,
\item the deviation of $\theta^\text{PMNS}_{23}$ from maximal mixing,
\item the neutrino mass ordering (normal or inverse),
\item the neutrino mass scale,
\item whether neutrino masses are of Dirac- or Majorana-type,
\item and if neutrinos have Majorana masses, the values of the Majorana CP phases.
\end{itemize}

\section{The Neutrino Puzzles}

There are two main puzzles in the neutrino sector: The first one may be called the ``mass puzzle''. It summarises the challenge to find the right extension of the Standard Model (SM) giving rise to the observed neutrino masses and to explain their smallness. The second one may be called the ``neutrino flavour puzzle'', containing the question of why the leptonic mixing angles are large compared to the quark mixing angles of the CKM matrix, and whether there is any pattern hidden in the leptonic mixing angles that could guide us towards the theory of flavour (where we still have a long way to go). 

\subsection{The Mass Puzzle}
As you all know, various mechanisms have been proposed to introduce masses for the neutrinos. For example, there are the well-known tree-level possibilites, the seesaw mechanisms (of type I \cite{typeI}, type II \cite{typeII} and type III \cite{typeIII}). But there are also proposals where, for example, neutrino masses emerge only at loop level, or from non-perturbative effects in string theory. 

Many models leading to Majorana neutrino masses can, at low energies relevant for neutrino oscillation physics, be described by the SM plus a dimension five neutrino mass operator \cite{Weinberg:1979sa} and eventually plus higher-dimensional operators leading, e.g., to a non-unitarity of the leptonic mixing matrix \cite{Belen}. However, neutrinos may also be Dirac particles, with the smallness of the masses explained, e.g., in extra-dimensional theories from a small overlap of the neutrino wave-functions with the four-dimensional brane where the SM fields are confined to \cite{Dienes:1998sb,ArkaniHamed:1998vp}. And finally, it may well be that the true mechanism realized in nature has not been proposed yet.

The present status is that there are no experimental hints to tell us which of the proposed mechanisms (if any) is the right one. It is not even known at which scale neutrino masses are generated. While many models based on a seesaw mechanism assume neutrino mass generation at very high energies (close to the GUT scale $M_\text{GUT} \sim 10^{16}$ GeV), the small neutrino masses can also be generated at energies around the TeV scale, or even at much lower energies such that light sterile neutrinos can propagate in oscillation experiments. 

To make progress towards this question, it will be necessary to combine all experimental data which will be available in the future, i.e.\ from the LHC, from indirect tests (for example from charged lepton flavour violation searches and tests of the non-unitarity of $U_\text{PMNS}$) as well as from experiments on the type and value of the absolute neutrino mass scale (from cosmology, Tritium $\beta$-decay, 0$\nu\beta\beta$-decay) and from neutrino oscillation experiments.

\subsection{The Neutrino Flavour Puzzle}

The ``neutrino flavour puzzle'', i.e.\ the puzzle of why the leptonic mixing angles are large compared to the quark mixing angles of the CKM matrix, and whether there is any pattern hidden in the leptonic mixing angles, will be the main part of my talk. Let us start by briefly reviewing the present knowledge of quark and lepton masses and mixings, which is summarized in table \ref{tab:masses_mixings}. 
Regarding the mixing angles, it is indeed striking that they are large in the lepton sector, while the largest mixing angle in the quark sector, the Cabibbo angle $\theta_C \equiv \theta_{12}^\text{CKM} \approx 13.0^\circ $, is of the same order as the recently measured smallest leptonic mixing angle $\theta_{13}^\text{PMNS} \approx 9^\circ \pm 1^\circ$. 
CP violation in the CKM matrix is parameterized by a large Dirac CP phase, $\delta^\text{CKM} \approx 69^\circ \pm 3^\circ $, while the CP phase(s) in the neutrino sector are still unknown.

\begin{table}

\begin{center}
Quark and lepton mixing angles:\\[1mm]
	\begin{tabular}{|l||l|}
	\hline
	$\theta_{12}^\text{PMNS} \approx 34^\circ \pm 1^\circ$   &  $\theta_{12}^\text{CKM} \approx 13.0^\circ $ \vphantom{$\frac{f}{f}$} \\
	$\theta_{23}^\text{PMNS} \approx 46^\circ \pm 3^\circ$   &  $\theta_{23}^\text{CKM} \approx 2.4^\circ $ \\
	$\theta_{13}^\text{PMNS} \approx 9^\circ \pm 1^\circ$     &  $\theta_{13}^\text{CKM} \approx 0.2^\circ $ \\
	$\delta^\text{PMNS}$ unknown     &  $\delta^\text{CKM} \approx 69^\circ \pm 3^\circ $ \\
	\hline
	\end{tabular}\\[3mm]
	
Quark masses (at $\mu = m_t$):\\[1mm]
	\begin{tabular}{|l|l|}
	\hline
	$m_u \approx 0.0012$ GeV  & $m_d \approx 0.0028$ GeV  \\
	$m_c \approx 0.590$ GeV  & $m_s \approx 0.052$ GeV    \\
	$m_t \approx 162.9$ GeV  & $m_b \approx 2.8$ GeV  \\
	\hline
	\end{tabular}\\[3mm]
	
Lepton masses:\\[1mm]
	\begin{tabular}{|c|l|}
	\hline
	 $|m_3^2 - m_1^2| \approx 2.4 \times 10^{-3}$ eV$^2$   & $m_e \approx 0.0005$ GeV \\
	 $\:m_2^2 - m_1^2   \approx 7.6 \times 10^{-5}$ eV$^2$   &   $m_\mu  \approx 0.0102$ GeV \\
	 $m_i < {\cal O}(0.5\:\text{eV})$      & $m_\tau \approx 1.8$ GeV \\
	\hline
	\end{tabular}
		
	\caption{Overview over the present knowledge of quark and lepton masses. For quarks and charged leptons, the running masses at the top mass scale, $\mu = m_t (m_t)$ are given \cite{Xing:2007fb}. The neutrino results are based on the global fit from \cite{Schwetz:2011zk}. Hints for $\theta_{23}^\text{PMNS}<45^\circ$ have been reported in \cite{Fogli:2011qn}.}
	\label{tab:masses_mixings}
	
\end{center}
\end{table}

Regarding the masses, we see that in the quark and charged lepton sectors, there is a strong hierarchy between the masses of the three families. This hierarchy is strongest in the up-type quark sector, while the masses are less hierarchical in the down-type quark and charged lepton sectors, and similar for each family of down-type quarks and charged leptons. This similarity may be a consequence of Grand Unified Theories (GUTs). We will come back to this possibility in the context of models for large $\theta^\text{PMNS}_{13}$ later in my talk. Of course in the neutrino sector only two mass splittings are known and the absolute neutrino mass scale is only constrained to be (roughly) below 0.5 eV from cosmology, Tritium $\beta$-decay and 0$\nu\beta\beta$-decay experiments.

\section{Ideas and Approaches}

\subsection{Top-down Approaches}
There are various ways to approach the above-mentioned neutrino puzzles. Possible top-down viewpoints are, for example:

\begin{itemize}

\item {\bf Anarchy:} In anarchy \cite{Hall:1999sn}, the idea is that the large mixing in the lepton sector is the result of a lack of structure. When the neutrino mass matrix is filled with random entries, it is indeed plausible that the resulting mixing angles are large. In fact, if $\theta_{13}^\text{PMNS}$ would have been very small, this would have been an argument against this viewpoint. With the now measured not so small value, this idea is still alive. However, it is also hard to test since only probabilities are predicted, but not specific values of parameters, which can be tested in experiments. 

\item {\bf Family symmetries:} Here the approach is to explain the neutrino mixing properties by assuming additional structure, generated by additional symmetries beyond the gauge symmetries of the SM. To generate flavour structure, such symmetries distinguish between members of different families. They are therefore called family symmetries or horizontal symmetries. Such symmetries can, on the one hand, explain the hierarchy of the masses in the quark and charged lepton sectors \cite{Froggatt:1978nt}, but they can also explain the large mixing in the lepton sector. Non-Abelian discrete symmetry groups, like e.g.\ $A_4$ or $S_4$, have become very popular in this respect \cite{discrete_models}.   

\item {\bf Grand Unified Theories (GUTs):} In GUTs \cite{Georgi,SO(10)}, the forces of the SM can be unified at high energies, around $M_\text{GUT} = 2 \times 10^{16}$ GeV. Left-right-symmetric GUTs are also appealing because they predict small neutrino masses via a version of the seesaw mechanism. In general, GUTs not only unify the forces of the SM, but also different types of particles are unified in joint representations of the GUT symmetry group. This GUT symmetry group contains the SM gauge group as a subgroup and, compared to family symmetries, it acts ``vertically''. This means it is linking properties of different particle types, such that the flavour structures of quarks and leptons are no longer disconnected. 

\item {\bf Extra Dimensions and String Theory:} Alternatively, one may also attempt to approach the neutrino puzzles directly from extra-dimensional theories or from the perspective of string theory. As already mentioned, one approach to explain the smallness of Dirac-type neutrino masses uses a small overlap of the neutrinos wave-function with our four-dimensional brane \cite{Dienes:1998sb,ArkaniHamed:1998vp}. In another approach, for instance, neutrino masses and large lepton mixing emerge from string theory instanton effects \cite{Blumenhagen:2006xt,Antusch:2007jd}.  

\end{itemize}

\subsection{Bottom-up Suggestions}
In addition, there have been various suggestions for relations between mixing parameters, inspired by bottom-up considerations based on the improved accuracy of the experimental data. Suggested relations of this type are, e.g.:

\begin{itemize}

\item {\bf Tri-bimaximal mixing:} Tri-bimaximal (TB) mixing \cite{tribi} is a specific mixing pattern, defined by
\be \label{eq:TM_mixing_angles}
\theta_{12}^\text{PMNS} = \arcsin \left( \frac{1}{\sqrt{3}}\right), \;\theta_{23}^\text{PMNS} = 45^\circ ,\;  \theta_{13}^\text{PMNS} = 0^\circ \,.
\ee
Before the measurement of $\theta_{13}^\text{PMNS}$, it was in excellent agreement with the experimental data. With its prediction $\theta_{13}^\text{PMNS} = 0^\circ$, it is now known that TB mixing can not hold exactly. However, as we will discuss later, it may still be a viable leading order structure of the neutrino mixing matrix.  

\item {\bf Bimaximal mixing:} The bimaximal (BM) mixing pattern \cite{Barger:1998ta} was proposed earlier than TB mixing, and is defined by 
\be
\theta_{12}^\text{PMNS} = 45^\circ, \;\theta_{23}^\text{PMNS} = 45^\circ ,\;  \theta_{13}^\text{PMNS} = 0^\circ \,.
\ee
Of course, $\theta_{13}^\text{PMNS}$ as well as $\theta_{12}^\text{PMNS}$ have to be corrected, compared to the BM prediction, in order to be consistent with the present experimental data. 

\item {\bf Quark-lepton complementarity:} Quark-lepton complementarity (QLC) suggests a link between the quark and lepton mixing angles of the form
\be
\theta_{12}^\text{PMNS} + \theta_C = 45^\circ \;,
\ee
which is called the QLC relation \cite{QLC}. It was discussed that this relation (and other similar relations) may be obtained by multiplying a bi-maximal mixing matrix times a mixing matrix equal to the CKM matrix \cite{QLC,Minakata:2004xt}.


\item Finally, in the light of the recent measurement of $\theta^{\text{PMNS}}_{13} = 8.8^\circ \pm 1.0^\circ$, which is consistent with 
\be \label{Eq:3}
\theta^{\text{PMNS}}_{13} =  \frac{\theta_C}{\sqrt{2}}    \:,
\ee
it has been suggested that this value of $\theta^{\text{PMNS}}_{13}$ 
may be a footprint of an underlying GUT, and it has been discussed that it can indeed emerge as a (leading order) {\bf prediction in realistic GUT models under simple conditions} \cite{Antusch:2012fb}.\footnote{In the context of some versions of QLC, where a bi-maximal mixing matrix is multiplied with a matrix equal to the CKM matrix in a certain way, the relation of Eq.~(\ref{Eq:3}) has been mentioned in \cite{Minakata:2004xt}. Recently, a modification of the TB mixing scheme with this value of $\theta^{\text{PMNS}}_{13}$ (with unchanged $\theta^{\text{PMNS}}_{12}$ and $\theta^{\text{PMNS}}_{23}$) has been suggested in \cite{King:2012vj}.}

\end{itemize}

One main theme in the last years has been to try to connect bottom-up suggestions to model building approaches following the various top-down viewpoints.  In my talk, I will now discuss two examples, and I apologize if I can not cover your favourite topic (due to lack of time, respectively space).

\section{Family Symmetries and Mixing Patterns}

One aspect, which received a lot of attention in the last years, is the possibility to realize specific neutrino mixing patterns, in particular TB mixing, with the help of family symmetries like, for instance, $A_4$ and $S_4$. In model-building, one may distinguish two possibilities, which may be called (following \cite{King:2009ap}) direct models and indirect models. For the following discussion we will focus on the example of TB mixing, where the leptonic mixing matrix takes the special form \cite{tribi}
\begin{eqnarray}
U_\text{TB} =
\left( \begin{array}{rrr}
\sqrt{\frac{2}{3}}  & \frac{1}{\sqrt{3}} & 0 \\
-\frac{1}{\sqrt{6}}  & \frac{1}{\sqrt{3}} & \frac{1}{\sqrt{2}} \\
\frac{1}{\sqrt{6}}  & -\frac{1}{\sqrt{3}} & \frac{1}{\sqrt{2}}
\end{array}
\right).
\label{TBmixing}
\end{eqnarray}   
leading to the TB pattern of mixing angles in Eq.~(\ref{eq:TM_mixing_angles}).

\subsection{Preliminary Remark: Symmetries which enforce TB Mixing are Broken Symmetries}

How can family symmetries be related to mixing patterns like TB mixing? The first idea could be that there might be a symmetry which one just has to impose on the theory (as an unbroken exact symmetry) and which would then force the mixing to TB form. One can easily see that this is not possible. 

If such a symmetry would exist for the whole theory, then one consequence would be that the mixing pattern is stable under renormalization group (RG) running. However, one can show that this is not the case. It therefore follows that no unbroken symmetry can exist to enforce TB mixing. Of course this is not the end of the story, as I will explain in the next two subsections, however one should keep in mind that we are talking about broken family symmetries.   

Let us discuss the RG running effects in a bit more detail: Below the seesaw scales, i.e.\ the masses of the right-handed neutrinos in type I seesaw models, and up to ${\mathcal{O}} (\theta_{13})$ corrections, the evolution of the
mixing angles is given by \cite{Antusch:2003kp} (dropping here the PMNS labels for the mixing parameters for brevity)
\begin{eqnarray}
\label{Eq:t12}\dot{\theta}_{12} \!&=&\!
        \frac{- C_e y_\tau^2}{32\pi^2} \,
        \!\sin 2\theta_{12} \, s_{23}^2\,  
      \frac{
      | {m_1}\, e^{i \varphi_1} \!\!+\! {m_2}\, e^{i  \varphi_2}|^2
     }{\Delta m^2_{21} },\nonumber\\
     &&
\\ \nonumber
\label{Eq:t13}\dot{\theta}_{13} \!&=&\!
        \frac{C_e y_\tau^2}{32\pi^2} \,
        \sin 2\theta_{12} \, \sin 2\theta_{23} \,
        \frac{m_3}{\Delta m^2_{31} \left( 1+\zeta \right)} \nonumber \\
      &&  \times \: I(m_i, \varphi_i, \delta)
\,,
\\ \nonumber
\label{Eq:t23}     \dot{\theta}_{23} \!&=&\! 
        \frac{-C_e y_\tau^2}{32\pi^2} \,
        \frac{\sin 2\theta_{23}}{\Delta m^2_{31}}
       \left[
         c_{12}^2 \, |m_2\, e^{i \varphi_2} + m_3|^2   \right.      \nonumber \\
      && + \!\left. s_{12}^2 \, \frac{|m_1\, e^{i \varphi_1}\! + m_3|^2}{1+\zeta}
        \right].
\label{eq:Theta23Dot}
\end{eqnarray}
The dot indicates differentiation $d/dt \equiv \mu\,d/d\mu$ (with $\mu$ being
the renormalization scale), and the abreviations $s_{ij}=\sin\theta_{ij}$,  
$c_{ij} = \cos\theta_{ij}$, $\zeta = \Delta m^2_{21}/\Delta m^2_{31}$ (with the mass squared differences $\Delta m^2_{ij} = m_i^2 - m_j^2$) 
and 
\begin{eqnarray}
\lefteqn{I(m_i, \varphi_i,\delta)
=\left[ m_1 \cos(\varphi_1-\delta) \right.}\nonumber \\
&& \left.  
- ( 1\!+\!\zeta ) m_2 
\cos(\varphi_2-\delta) - \zeta m_3  \cos\delta \right]
\end{eqnarray} 
have been used. In the SM, $C_e = -3/2$, while in the MSSM, $C_e = 1$. $y_\tau$ denotes the tau Yukawa coupling, and one can safely neglect the contributions coming from the electron and muon Yukawa couplings. For the Majorana phases $\varphi_1$ and $\varphi_2$, we use the same convention  as in \cite{Antusch:2003kp}. 

From these expressions one can easily estimate the typical size of RG effects on the leptonic mixing angles and see some basic properties. In particular, one can see that TB mixing is not stable under RG running.\footnote{Analytical formulae including the effects of the neutrino Yukawa couplings, relevant above the mass thresholds of the right-handed neutrinos, can be found in \cite{Antusch:2005gp}.}

\subsection{TB Mixing in Direct Models}

Despite the fact that family symmetries for TB mixing have to be broken, there is still hope to identify useful underlying family symmetries and to apply them for realizing the TB  mixing pattern. In direct models, the idea is to make use of remnant symmetries in the neutrino and the charged lepton sectors. If an underlying family symmetry group $G_F$ had existed (and was then spontaneously broken) such remnant symmetries might have survived, and by identifying them one could at least recover part of $G_F$. Such considerations lead to the family symmetry group $S_4$ \cite{Lam:2008sh} or closely related groups as the minimal possibilities for $G_F$.\footnote{We note that also $A_4$ is suitable for constructing direct models with TB mixing, if only part of the available group representations are used.}

For instance, assuming TB mixing and working in the neutrino flavour basis, one finds that the generators $S$ and $U$ (for definitions, see, e.g., \cite{King:2009ap}) are preserved in the neutrino sector, while the diagonal generator $T$ is preserved in the charged lepton sector. Combining them one arrives at the symmetry group $S_4$. Now engineering backwards, one can make sure that a model generates TB mixing if one first postulates a symmetry group $G_F$ containing $S_4$ and then breaks it spontaneously such that the symmetries generated by $S$, $U$ respectively $T$ survive in the neutrino and charged lepton sectors as remnant symmetries.

\subsection{TB Mixing in Indirect Models}

While in direct models the symmetry group plays the central role, the key feature of indirect models are the directions in flavour space in which the family symmetry is broken. In fact, if the symmetry group $G_F$ is broken along the directions given by the columns of the TB mixing matrix $U_\text{TB} $ of Eq.~(\ref{TBmixing}), then TB mixing in the neutrino sector can easily be realised, independent of the neutrino mass eigenvalues. This can be used in model building by promoting the columns of $U_\text{TB} $ to dynamical fields, the so-called flavons, and by making their vacuum expectation values point in the desired directions in flavour space.\footnote{We note that this concept may easily be applied in seesaw models, but it can also be used in other scenarios, e.g.\ in the context of string theory models where the neutrino masses emerge from instanton effects \cite{Antusch:2007jd}. In seesaw scenarios, it is called form dominance \cite{Chen:2009um}.}    

Although here the choice of the symmetry group is not as crucial as in direct models, it has turned out that for making the flavons align in the right directions, non-Abelian discrete symmetries (like $A_4$ or $S_4$) are favourable. In the recent years, various models have been constructed by many authors following the direct or indirect approach, and using non-Abelian discrete groups as family symmetry $G_F$.

\subsection{TB Mixing Confronted with Recent $\theta^{\text{PMNS}}_{13}$ Results}

As we have already stated above, exact TB mixing (which features zero $\theta^{\text{PMNS}}_{13}$) is ruled out by the latest experimental results for $\theta^{\text{PMNS}}_{13}$. Model builders show two possible reactions w.r.t.\ this fact:

\begin{itemize}

\item One possible reaction is to {\bf give up TB mixing and to look for alternative structures} which already feature non-zero large $\theta^{\text{PMNS}}_{13}$. Within the direct approach to model building, this could for example imply to use different discrete symmetry groups, as e.g.\ in \cite{deAdelhartToorop:2011re}. In the context of indirect models, such structures can arise from breaking the familly group $G_F$ along a new direction in flavour space, as e.g.\ in CSD2 \cite{Antusch:2011ic} which leads to so-called tri-maximal mixing with predicted value of $\theta^{\text{PMNS}}_{13}$.

\item Another possible reaction is to {\bf view TB mixing as a leading order pattern only}, and to apply corrections to it. Such corrections could, for example, be applied to the family symmetry breaking vacuum expectation values of the flavon fields \cite{King:2011ab}. Another, quite popular and in fact very well motivated correction is provided by charged lepton mixing contributions (see e.g.~\cite{Antusch:2005kw}), as we will discuss in more detail in the following section.

\end{itemize}

\section{$\theta^{\text{PMNS}}_{13}$ from charged lepton mixing effects - are there GUT footprints in the PMNS matrix?}

\begin{figure}
 \center
 \begin{tikzpicture}[
      scale=0.45,
      circ/.style={thick,circle},
      rec/.style={thick,rectangle},
      trans/.style={thick,<->,shorten >=2pt,shorten <=2pt,>=stealth},
    ]
    \draw[circ] (0,8) node[style={fill=black!20},scale=1.2] {$Y_u$};
    \draw[circ] (8,8) node[style={fill=black!20},scale=1.2] {$Y_d$};
    \draw[circ] (8,0) node[style={fill=black!20},scale=1.2] {$Y_e$};
    \draw[circ] (0,0) node[style={fill=black!20},scale=1.2] {$m_\nu$};

    \draw[trans] (2,0) -- (6,0) node[midway,below,scale=1] {$U_\text{PMNS} = U^{e\dagger} U^\nu$};
    \draw[trans] (2,8) -- (6,8) node[midway,below,scale=1] {$U_\text{CKM} = U^{u\dagger} U^d$};
    \draw[trans] (8,6) -- (8,2) node[midway,below,scale=1,rotate=90] {GUT relations};
    \end{tikzpicture}
\caption{Using GUT relations between the down-type quark and charged lepton Yukawa matrices, and some simple conditions, charged lepton mixing effects can induce $\theta_{13}^{\text{PMNS}} \approx \theta_C / \sqrt{2}$ in GUT models.}
\label{picture}
\end{figure}
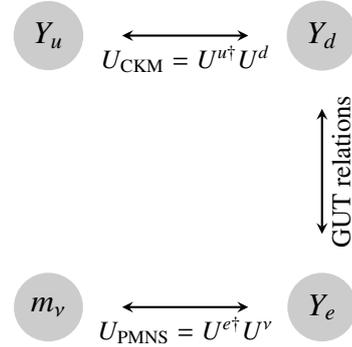

In GUT models of flavour, non-zero $\theta^{\text{PMNS}}_{13}$ is generically expected due to the presence of charged lepton mixing contributions.  The general picture is illustrated in Fig.~\ref{picture}. The quark mixing matrix is composed from the left diagonalization matrices of the down-type and up-type quark mass matrices (or of the corresponding Yukawa matrices $Y_d$ and $Y_u$), $U_\text{CKM} = U^{u\dagger} U^d$, and analogously the lepton mixing matrix is composed from the left diagonalization matrices of the neutrino mass matrix $m_\nu$ and of the charged lepton Yukawa matrix $Y_e$, i.e.\ $U_\text{PMNS} = U^{e\dagger} U^\nu$. 

In GUTs, quarks and leptons are unified in joint representations of the GUT symmetry group, which implies that the elements of the quark and lepton Yukawa matrices can arise (dominantly) from single joint operators. As a consequence, the flavour structures of the quark and the lepton sectors are 
linked and $Y_d$ and $Y_e$ are connected via ``GUT relations''. Under some simple conditions, these GUT relations can lead to predictive schemes for $\theta^{\text{PMNS}}_{13}$, which may appear as ``footprints'' of GUTs in the PMNS matrix, as we will discuss below. One possible footprint is the phenomenologically attractive relation $\theta^{\text{PMNS}}_{13} \approx \theta_C / \sqrt{2}$.

\subsection{$\theta^{\text{PMNS}}_{13} \approx \theta_C / \sqrt{2}$ from GUTs}

Let us now discuss, following \cite{Antusch:2012fb}, how predictions for $\theta^{\text{PMNS}}_{13}$, linked to the Cabibbo angle of the quark mixing matrix, can arise in GUT models:

\begin{itemize}

\item Our starting point is the neutrino mass matrix. Let us consider the case that the 1-3 mixing in the neutrino sector (here referred to as $\theta_{13}^\nu$) is negligibly small,
\be
\theta_{13}^\nu \approx 0 \;,
\ee
as it holds true, for example, in TB neutrino mixing and bimaximal neutrino mixing. 

\item Turning to the quark sector, we can see from Tab.~\ref{tab:masses_mixings} that the hierarchy between the masses of the three families is stronger in the up-type quark sector than in the down-type quark sector. With hierarchical mass matrices, the stronger hierarchy is generically associated with the smaller mixing, such that we often encounter
$\theta_{ij}^u \ll \theta_{ij}^d$, leading to
\be\label{eq:t12d_tC}
\theta_{12}^d \approx \theta_C \;,
\ee
while the other mixing angles in $Y_d$ are much smaller than the Cabibbo angle and will be neglected in the following discussion.

\item As mentioned above, GUT relations connect quark and lepton masses, when the elements of the quark and lepton Yukawa matrices arise dominantly from single joint GUT operators. In an analogous fashion, this also leads to relations between quark and lepton mixing angles. In particular, the 1-2 mixings in $Y_e$ and $Y_d$ are then connected by ratios of Clebsch factors $c_{ij}$, for example by \cite{Antusch:2011qg,Marzocca:2011dh}\footnote{Available Clebsch factors in SU(5) GUTs are, e.g., $ \{\frac{1}{2},1,\frac{3}{2},3,\frac{9}{2},6,9\}$ and in Pati-Salam models, e.g., $ \{\frac{3}{4},1,2,3,9\}$. For the corresponding GUT operators and their viability in supersymmetric scenarios, see e.g.~\cite{Antusch:2009gu}.} 
\be\label{eq:t12e_tC}
\theta_{12}^e \approx \frac{c_{12}}{c_{22}} \theta_{12}^d \approx \frac{c_{12}}{c_{22}} \theta_C \;,
\ee  
where Eq.~(\ref{eq:t12d_tC}) has been used in the last step.

\item Finally, recalling that $\theta_{13}^\nu \approx 0$ (and that also $\theta_{13}^e$ is negligibly small), one finds that the charged lepton mixing contribution $\theta_{12}^e$ (via $U_\text{PMNS} = U^{e\dagger} U^\nu$) now determines $\theta^{\text{PMNS}}_{13}$. In addition, $\theta_{12}^e$ also changes the solar mixing angle $\theta^{\text{PMNS}}_{12}$, as we will discuss in the next subsection. The induced $\theta^{\text{PMNS}}_{13}$ is given by 
\be\label{Eq:t13fromt12e}
\theta^{\text{PMNS}}_{13} \approx \theta_{12}^{e}s_{23}^\text{PMNS}  \approx  \frac{\theta_{12}^{e}}{\sqrt{2}}  \approx \frac{\theta_C}{\sqrt{2}}  \frac{c_{12}}{c_{22}}\;,
\ee
where in the second step a maximal mixing $\theta^{\text{PMNS}}_{23}$ has been plugged in, and in the third step the ''GUT mixing relation'' Eq.~(\ref{eq:t12e_tC}) has been used. One can see that $\theta^{\text{PMNS}}_{13}$ is predicted in terms of the Cabibbo angle times a ratio of Clebsch factors.\footnote{It should be noted in this context that many existing GUT models have predicted smaller values of $\theta^{\text{PMNS}}_{13}$ (often around $3^\circ$), especially when they were using the Georgi-Jarlskog Clebsch factor of $3$ to obtain viable mass relations for the second and the first quark and lepton families \cite{GJ}.} 
The specific relation $\theta^{\text{PMNS}}_{13} \approx \theta_C / \sqrt{2}$ arises when two Clebsch factors are set equal.\footnote{We note that the conditions are somewhat different in SU(5) GUTs and in Pati-Salam unified models. Details are left out here for brevity and can be found in \cite{Antusch:2012fb}.}

\end{itemize}

\subsection{The  Lepton Mixing Sum Rule}

In the scenario described in the previous subsection, the charged lepton mixing contribution $\theta_{13}^e$ also modifies the 1-2 mixing of $U^\nu$ according to 
$\theta_{12}^{\text{PMNS}}  \approx \theta_{12}^\nu + \frac{\theta_{12}^e}{\sqrt{2}} \cos (\delta^{\text{PMNS}} )$, assuming maximal $\theta^{\text{PMNS}}_{23}$ here for simplicity. 
Using Eq.~(\ref{Eq:t13fromt12e}), one obtains the following relation, 
\be\label{Eq:SumRule}
\theta_{12}^{\text{PMNS}} - \theta^{\text{PMNS}}_{13} \cos (\delta^{\text{PMNS}} ) \approx \theta_{12}^\nu\; ,
\ee 
known as the lepton mixing sum rule \cite{Antusch:2005kw,sumrule2}. 

\begin{figure}
\center
\includegraphics[scale=.455]{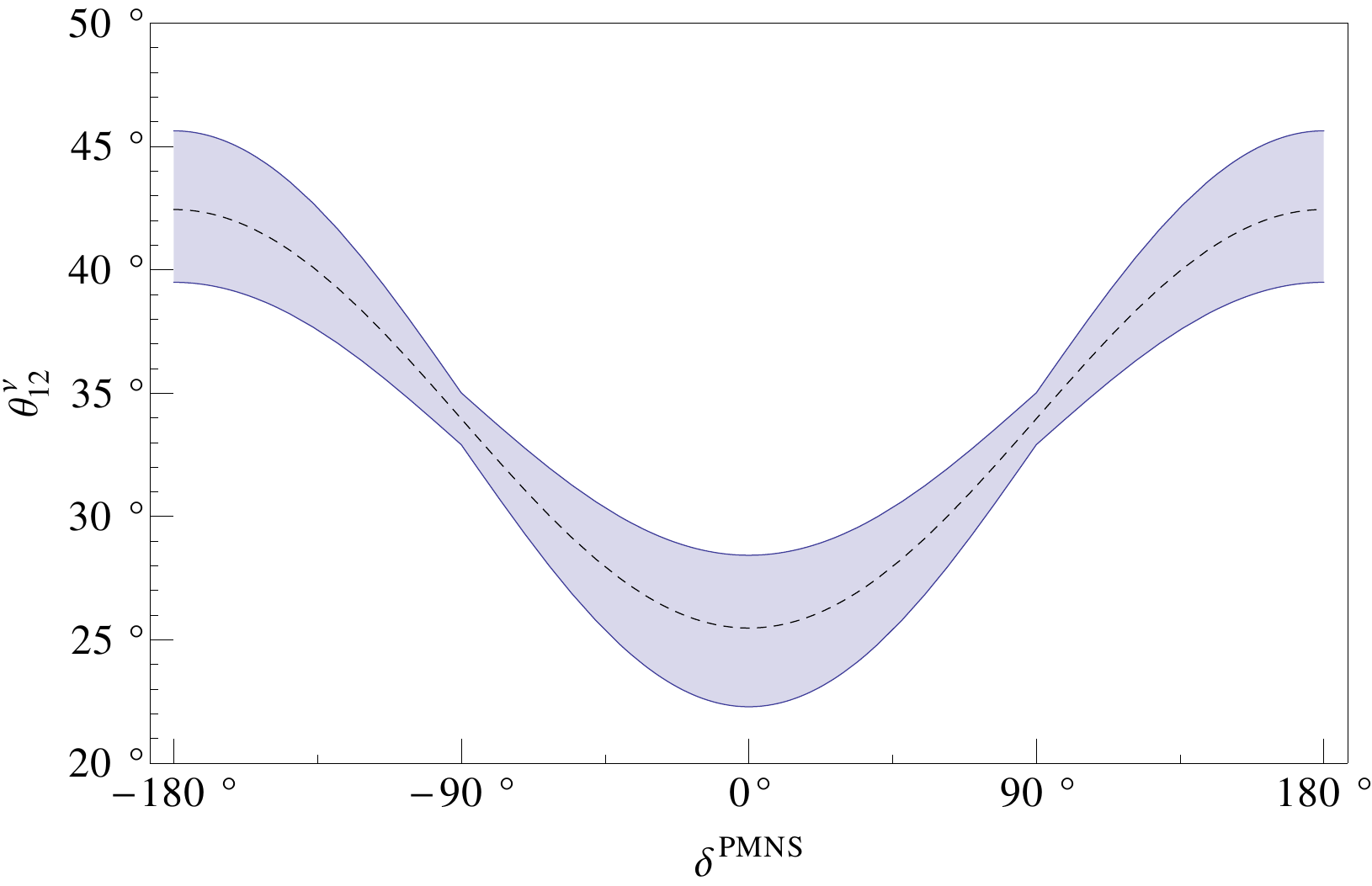}
\caption{Using the lepton mixing sum rule of Eq.\ (\ref{Eq:SumRule}), a measurement of $\delta^\text{PMNS}$ allows to reconstruct $\theta_{12}^\nu$ (provided that $\theta_{13}^\nu$, $\theta_{13}^e\ll\theta_C$). The shaded region corresponds to the present $1\sigma$ uncertainties for $\theta_{12}^\text{PMNS}$, $\theta_{13}^\text{PMNS}$ and $\theta_{23}^\text{PMNS}$, using the version of the sum rule with general $\theta_{23}^\text{PMNS}$ \cite{Antusch:2005kw,sumrule2}. The figure is taken from \cite{Antusch:2012fb}.}
\label{figuresumrule}
\end{figure}

With a future measurement of $\delta^{\text{PMNS}}$, the lepton mixing sum rule can be used to test whether a special mixing pattern, like e.g.\ TB mixing or bimaximal mixing, is realized in the neutrino sector. This is illustrated in Fig.~\ref{figuresumrule}. TB neutrino mixing, for example, would correspond to $\theta_{12}^\nu = \arcsin(1/\sqrt{3}) \approx 35.3^\circ$ and can only be viable for $\delta^{\text{PMNS}} \approx \pm 90^\circ$, whereas bimaximal neutrino mixing with $\theta_{12}^\nu =  45^\circ$ requires $\delta^{\text{PMNS}} \approx 180^\circ$.\footnote{Such specific predictions for $\delta^{\text{PMNS}}$ may emerge from flavour models with $\mathbb{Z}_2$ or $\mathbb{Z}_4$ shaping symmetries to explain a right-angled \text{\text{CKM}} unitarity triangle (with $\alpha \approx 90^\circ$), as discussed recently in \cite{Antusch:2011sx}.}

\section{Summary and Conclusions}

The recent measurement of $\theta^{\text{PMNS}}_{13} = 8.8^\circ \pm 1.0^\circ$ is exciting news for model building.  A large fraction of the proposed models are now ruled out, and new ideas are being developed towards explaining the experimentally found value.

While exact mixing patterns with vanishing $\theta^{\text{PMNS}}_{13}$, like for example TB mixing of $U_\text{PMNS}$, are ruled out, they may still provide viable leading order structures. 
Especially when mixing patterns are realized in the neutrino sector, with $\theta^{\text{PMNS}}_{13}$ originating from charged lepton 1-2 mixing contributions, this can lead to alternative scenarios with high predictivity. 

Independent of such neutrino mixing patterns, it is interesting that the phenomenologically attractive relation $\theta^{\text{PMNS}}_{13} \approx \theta_C / \sqrt{2}$ can be a consequence of an underlying Grand Unified Theory.

After this great experimental success of measuring $\theta^{\text{PMNS}}_{13}$, the next goals will include the measurement of the neutrino mass ordering, the Dirac CP phase $\delta^{\text{PMNS}} $, and the deviation of $\theta^{\text{PMNS}}_{23}$ from maximal mixing. In addition, $\theta^{\text{PMNS}}_{13}$ will be measured even more precisely, with an accuracy goal of about $\pm 0.25^\circ$. The results of these measurements will provide the next crucial input for model building.

Finally, I would like to note that given the (already present) high experimental precision for the leptonic mixing parameters, it is important that also the theoretical analyses of models for neutrino masses and mixings are performed with comparable accuracy.

\section*{Acknowledgments}
I would like to thank the organizers of Neutrino 2012 for their effort to make this great conference possible. Furthermore, I want to thank the Swiss National Science Foundation for support.


\end{document}